\documentclass[twoside]{LCWS11}
\usepackage[latin1]{inputenc}
\usepackage[dvips]{graphicx,epsfig,color}
\usepackage{wrapfig,rotating}
\usepackage{amssymb,amsmath,array}
\usepackage{cite}
\begin{document}
\title{
Improving the Prompt Electromagnetic 
Energy Component of Jet
Energy Resolution with $\pi^{0}$
Fitting in High Granularity Electromagnetic Calorimeters} 
\author{Brian van Doren and Graham W. Wilson
\thanks{Supported in part by the U.S. National Science Foundation under grant NSF-PHY-0935539.}
\vspace{.3cm}\\
University of Kansas, Dept. of Physics and Astronomy, Lawrence, KS 66045, USA
}

\maketitle

\begin{abstract}
We investigate improving the hadronic jet energy resolution using 
mass-constrained fits of $\pi^{0} \rightarrow \gamma \gamma$ decays using high 
granularity electromagnetic calorimeters. 
Single $\pi^{0}$ studies have indicated a large potential 
for improvement in the energy resolution of $\pi^0$'s, 
typically reducing the average energy resolution 
by a factor of two for 4 GeV $\pi^0$'s.
We apply this method to fully simulated multi-hadronic events 
with multiple $\pi^{0}$'s with widely varying energies 
using the ILD00 detector model.
Several methods for identifying the correct pairings of 
photons with parent $\pi^{0}$'s were explored. 
The combinatorics become challenging as the number 
of $\pi^{0}$'s increases and we employ the Blossom V implementation 
of Edmonds' matching algorithm for handling this.
For events where both photons of the $\pi^0$ are detected, 
the resulting solutions lead to an improvement 
in the $\pi^{0}$ component of the event energy resolution 
for 91.2 GeV $Z^{0}$ events 
from $18.0\%/\sqrt{E}$ to $13.9\%/\sqrt{E}$ using 
the ILD00 detector and its reconstruction algorithms.
This can be compared to a maximum potential improvement to $12.2\%/\sqrt{E}$ 
if all photon pairs are matched correctly using the current photon reconstruction.
\end{abstract}

\section{Introduction}



The concept of particle flow calorimetry~\cite{Brient, Morgunov} 
is widely seen as one of the most promising approaches for 
reconstructing the energy of hadronic jets in 
$\mathrm{e}^+ \mathrm{e}^-$ colliders. 
Much progress has been made on the 
particle-flow algorithms, particularly as described in~\cite{Thomson}.
The envisaged high granularity detectors open
up new avenues to explore and further improve the event 
reconstruction in the detector designs envisaged for 
future accelerator facilities such as the ILC 
and CLIC $\mathrm{e}^+ \mathrm{e}^-$ linear colliders.

The ability to reconstruct accurately individual photons in the midst 
of hadronic jets leads to the possibility 
to reconstruct individual di-photon resonances such as $\pi^0$'s.
A typical multi-hadronic event 
consists on average of about $62\%$ of the energy in charged hadrons, $28\%$ in photons, 
and $10\%$ in neutral hadrons with large event-to-event fluctuations. 
The photons are primarily from $\pi^{0} \rightarrow \gamma\gamma$ and 
we aim to reconstruct these intermediate $\pi^{0}$'s.
The results from mass-constrained fits can significantly improve 
the prompt electromagnetic component of jet energy resolution.

\section{Overview}
%
%
The focus of this paper is to investigate the performance 
improvements that 
may be obtained for the reconstruction of the energy of multiple $\pi^0$'s 
in multi-hadronic decays of the $Z^0$ using mass-constrained fits based on 
a realistic full simulation and event reconstruction of the ILD detector. 
We first briefly discuss mass-constrained fits in section~\ref{sec:fits} and then 
review some of the results obtained with toy Monte Carlo studies for both single $\pi^0$'s and 
multiple $\pi^0$'s from $Z^0$ decays in section~\ref{sec-toyMC}.
We then concentrate in section~\ref{sec-fullSIM} on the results obtained from 
full simulation of the ILD detector for events with multiple $\pi^0$'s from 
$Z^0$ decay at rest. A central issue is how to best 
pair up sibling photon candidates to parent $\pi^0$'s. 
We use three pairing algorithms:
\begin{enumerate}
\item{Cheated (using MC truth information)} 
\item{A realistic matching algorithm with 
no recourse to truth information using Blossom V}
\item{Same as 2, except that incorrect pairings in the solution are 
identified (using MC truth information) and replaced with 
the original photon measurements}
\end{enumerate}

We examine the mass-constrained fitting performance 
for the three different pairing algorithms and compare the results with 
the standard calorimetric measurement with no mass-constrained fits.
In order to make the study tractable with the 
current reconstruction software we have restricted  
to simulating only the prompt $\pi^0$ component 
of $Z^0$ decays (ignoring potential complications 
from additional photon candidates originating from charged and neutral hadrons) 
and requiring that none of the photons convert in the tracking volume.

\section{Mass-Constrained Fits}
\label{sec:fits}

Given a $\pi^{0} \rightarrow \gamma \gamma$ decay, 
we apply a mass-constrained fit where the invariant mass of the two photons is 
constrained to the $\pi^0$ mass. The adjusted photon momenta 
result in an improved energy and direction resolution for the $\pi^0$.
In the fit we define the photon 3-vectors using the measured energy, $E$, 
and spherical polar coordinates, $\theta$ and $\phi$ for the photon directions as 
estimated from the reconstructed cluster positions in 
the calorimeter and the assumption of the photons originating from 
the interaction point. The errors on the measurements are defined assuming that the 
errors on $E$, $\theta$ and $\phi$ are independent as is the case when 
the two photons are spatially resolved.
The mass constraint relates the energies of the two photons, $E_1$ and $E_2$, to the 
opening angle $\psi_{12}$ between the two photons:
\begin{equation}
m^2 = 4 E_1 E_2 \sin^2{(\psi_{12}/2)}
\label{eq:m1}
\end{equation}
The fitting has been implemented with the MarlinKinfit package~\cite{MarlinKinFit}.
In addition to improving the $\pi^0$ energy resolution when the correct photons are paired 
the procedure leads to a well-defined error estimate for the fitted $\pi^0$ energy 
which can potentially be used to better assign errors on the overall 
event reconstruction at the individual event level. 
However the latter has not been the focus of this paper.
%

\section{Toy Monte Carlo}
\label{sec-toyMC}

\subsection{Single $\pi^0$}

Single $\pi^0$ studies have demonstrated strong potential 
improvements in $\pi^0$ energy resolution~\cite{Graham_OldTalks} 
when the photon directions ($\theta$ and $\phi$) can be inferred 
with high precision. 
To illustrate this we show below the dependence of 
the $\pi^0$ energy resolution on 
the relevant factors: photon angular resolution, $\pi^0$ energy 
and $\cos{\theta^*}$, where $\theta^*$
is the center-of-momentum (CM) decay angle.
These particular estimates were obtained using a toy Monte Carlo  
which assumed energy independent angular resolutions with the 
$\phi$ uncertainty related to the $\theta$ uncertainty by $\sigma_{\phi} = \sigma_{\theta} / \sin{\theta}$.  

Table~\ref{tab:1pi0} shows as expected that the improvement in the stochastic parameter 
of the energy resolution, $\alpha$ where $\sigma_{E}/E = \alpha/\sqrt{E}$, 
depends a great deal on the CM decay angle. 
The most dramatic improvement being 
for symmetric decay ($\cos{\theta^*} = 0$) with almost no improvement at all for 
highly asymmetric decay. For each bin in $\cos{\theta^*}$ the distributions 
are rather Gaussian.
For a reference case of a 4 GeV $\pi^0$ with a photon 
fractional energy resolution, $\sigma_{E}/E$ 
of $16\%/\sqrt{E}$ with $E$ in GeV and a $\theta$ resolution 
of 1 mrad, the $\pi^0$ energy resolution\footnote{Estimated from 
the rms of the observed non-Gaussian distribution (consisting presumably of several Gaussians for different $\cos{\theta^*}$)} 
improves to $9.4\%/\sqrt{E}$ averaged over all CM decay angles 
while for symmetric decay the resolution improves to $4.2\%/\sqrt{E}$.
The precise measurement of the opening angle between the two photons, $\psi_{12}$, 
is the reason for the improvement in energy resolution, and this can be influenced by 
improving the relative error on $\psi_{12}$ - either by increasing the angle by 
decreasing the $\pi^0$ energy for the same $\cos{\theta^*}$ 
or further improving the measurement precision of the photon directions. 
The behavior, averaged over CM decay angles, on $\pi^0$ energy and 
angular resolution are also shown in the 
table. 


\begin{table}[h]
\begin{center}
\begin{tabular}{c c | c c | c c}
\multicolumn{2}{c}{Dependence on $\sigma_{\theta}$} &
\multicolumn{2}{c}{Dependence on $E_{\pi^0}$} &
\multicolumn{2}{c}{Dependence on $\cos{\theta^{*}}$} \\
\hline
  $\sigma_{\theta}$ (mrad) & $\alpha$ (\%) &
  $E_{\pi^0}$ (GeV) & $\alpha$ (\%) &
  $\cos{\theta^{*}}$ & $\alpha$ (\%) \\
  
  8         &  14.3  & 32    &  15.6 & 0.95   &  15.3 \\
  4         &  12.3  & 16    &  14.3 & 0.75   &  12.1 \\
  2         &  10.4  & 8     &  11.3 & 0.55   &  8.9 \\
  1         &  9.4   & 4     &  9.4 & 0.35   &  6.4 \\
  0.5       &  9.1   & 2     &  8.8 & 0.15   &  4.7 \\
  0.25      &  9.0   & 1     &  8.4 & 0.0 & 4.2 \\
  0.125     &  9.0   & 0.5   &  7.8 \\
            &        & 0.25  &  7.0 \\
\hline

\end{tabular}
\caption{Measured values of the stochastic 
energy resolution parameter, $\alpha$,  
estimated from the rms of the distribution of the stochastically 
scaled energy residual, 
$(E - E_{\mathrm{gen}}) / \sqrt{E_{\mathrm{gen}}}$, where 
$E$ is the fitted $\pi^0$ energy estimate and $E_{\mathrm{gen}}$ is the generator energy.
The table shows the dependence on the angular resolution, $\pi^0$ energy, and $\cos{\theta^*}$ respectively.
Unless otherwise specified the table 
assumes a $\pi^0$ energy of 4 GeV, $\sigma_{\theta} = 1$~mrad, 
and an intrinsic electromagnetic calorimeter resolution, $\alpha_{\mathrm{det}}$ set to 
16\%.}
\label{tab:1pi0}
\end{center}
\end{table}

\subsection{Multiple $\pi^0$'s: $Z^{0} \rightarrow q\overline{q} (q = u, d, s) $}

We then explored the impact of $\pi^{0}$ 
mass-constrained fits in $Z^{0}$ events with center of mass energy of 91.2 GeV with 
multiple $\pi^0$'s.
The events were simulated in the same fashion as the single particle studies 
using a well behaved energy uncertainty following the 
usual $ \sigma_{E}/E = \alpha / \sqrt{E}$ model and fixed angular resolutions.
There are some important things to note about $\pi^{0}$'s in a $Z^{0}$ event 
with decay to light quarks. 
The average number of $\pi^{0}$'s is $9.42 \pm 0.32$~\cite{PDG} with a 
median energy of about 2.4 GeV. 
The energy distribution is highly skewed to lower energies with a few high energy 
particles to compensate. To see the impact on the electromagnetic contribution to the 
overall event uncertainty, we extracted the $\pi^{0}$ information\footnote{For these studies we only 
retained prompt $\pi^0$'s produced within 10 cm of the interaction point which decayed to two photons 
(rejecting the 1.2\% of Dalitz decays).} 
from the events and 
considered the impact on the quantity $\alpha = \sigma_{E} / \sqrt{E}$ for $\pi^{0}$'s only.
For these studies we only 
considered prompt $\pi^0$'s produced within 10 cm of the interaction point, 
and we also only considered $\pi^0$'s which decayed to two photons (98.8\% branching fraction).
Several combinations of energy and angular uncertainties were studied and 
summarized in Table~\ref{tab:Z}. The general trends are as expected.
Improved angular resolution results in improved fitted energy 
resolution averaged over the randomized CM decay angles of the ensemble of $\pi^0$'s.

\begin{table}[h]
\begin{center}
\begin{tabular}{ c c c c }
\hline
   $\alpha_{\mathrm{det}}$ (\%) & $\sigma_{\theta}$ (mrad) & $\alpha_\mathrm{{fit}}$ (\%) & $\alpha_\mathrm{{fit}}/\alpha_\mathrm{{det}}$ \\
   \hline
   8   &  0.25  & 4.94 & 0.62 \\
   16  &  0.25  & 9.75 & 0.61 \\
   32  &  0.25  & 21.2 & 0.66 \\
   \hline
   8   &  0.5  & 5.31 & 0.66 \\
   16  &  0.5  & 10.2 & 0.64 \\
   32  &  0.5  & 21.7 & 0.68 \\
   \hline
   8   &  1.0  & 5.78 & 0.72 \\
   16  &  1.0  & 10.9 & 0.68 \\
   32  &  1.0  & 22.5 & 0.70 \\
   
\hline
   
\end{tabular}
\caption{Energy resolution improvement for the prompt electromagnetic energy 
contribution to 91.2 GeV $Z^{0}$'s. The resolution after fitting, $\alpha_{\mathrm{fit}}$, 
was estimated from the rms of the distribution of the stochastically scaled energy sum 
residual, $(E - E_{\mathrm{gen}}) / \sqrt{E_{\mathrm{gen}}}$.
}
\label{tab:Z}
\end{center}
\end{table}

\section{Full Simulation}
\label{sec-fullSIM}

Full simulation testing of the fitting procedure was performed using the standard 
distribution of ilcsoft v01-09 for the ILD
detector version ILD00 described in the ILD Letter of Intent~\cite{LOI} with 
a 3.5~T solenoidal field.
This uses Mokka to simulate with GEANT4 the detailed shower development of photons 
in the material of the detector, and the Marlin based framework 
for reconstructing the corresponding energy deposits. 
The most relevant features of this detector are a hermetic 
electromagnetic calorimeter composed of Tungsten absorber plates and 
Silicon readout pads with cell sizes of 5 mm $\times$ 5 mm. 
The calorimeter has 29 longitudinal layers with fine sampling every 0.6 $X_0$ 
for the front 20 layers and coarser 1.2 $X_0$ sampling for the back 9 layers.
The octagonal barrel part is located at a radius exceeding 1.85 m, and the endcaps are 
located 2.45 m longitudinally from the interaction point.
 
Modifications to algorithms included revising the cluster position estimates to use 
the standard weighted center of gravity method. 
It is anticipated that further improvements to reconstruction of cluster position using 
shower fitting may lead to significantly improved angular resolution as 
discussed in ~\cite{Graham_position_talk}.

\subsection{Photon Resolution Modeling}

Performance of the fitting procedure relies 
on reliable estimation of the resolution and correction of any biases. 
The fit variables are the energy and angular directions, therefore
we modeled the resolutions for $\sigma_{E}$, $\sigma_{\theta}$, and $\sigma_{\phi}$. 
Estimates for resolution were obtained by fully simulating and 
reconstructing single photons using a broad range of 
energies (0.2, 0.4, 0.8, 1.6, 3.2, 6.4, 12.8, 25.6 GeV).
In order to be considered in estimating the resolution, we required that 
the single photons do not convert before the calorimeter and that 
there is only one reconstructed cluster.
Energy resolution was set to depend on energy 
and be independent of angle: $\sigma_{E} / \sqrt{E} = A + B \log_{2}(E)$ 
with $E$ in GeV.
The end-cap and barrel regions were evaluated independently. 
For the barrel it was found that $A = 0.1591 \pm 0.0004$ 
and $B = 0.0053 \pm 0.0002$ with a $\chi^{2}/\mathrm{dof} = 8.6/6$. For 
the end-caps,
$A = 0.1468 \pm 0.0004$ and $B = 0.0049 \pm 0.0002$ with 
a $\chi^{2}/\mathrm{dof} = 7.8/6$.


The angular resolutions, $\sigma_{\theta}$ and $\sigma_{\phi}$, were determined separately 
in the barrel and end-cap regions taking into account their energy dependence.
It was found that the angular resolution dependence on $\theta$ could 
not quite be adequately described using simple low order polynomials. 
Therefore events were grouped into ten bins along the barrel, equally spaced in $0 < |z| < 2329$ mm
(we verified that the response is $z$-symmetric). 
The end-caps were also grouped into ten bins equally spaced radially in the  500 mm~$< \rho < 1843$ mm range.
The energies used in this sample included the same energies as above, as 
well as 0.1 GeV. 
This sample space of $E, \rho$ and $E, z$ was then used to estimate 
the angular resolutions by performing bi-linear interpolation between known values.

\begin{table}[h]
\begin{center}
\begin{tabular}{ c c c c c c }
\hline
   & Energy (GeV) & Center Barrel & End of Barrel & Outer End-Cap & Inner End-Cap \\
   \hline
   $\sigma_{\theta}$ &  0.1  &  $ 1.25 \pm 0.03 $    & $ 0.87 \pm 0.03 $   &  $ 0.81 \pm 0.03 $  & $ 0.94 \pm 0.03 $ \\
   $\sigma_{\phi}$   &  0.1  &  $ 1.35 \pm 0.03 $    & $ 1.21 \pm 0.06 $   &  $ 1.21 \pm 0.06 $  & $ 3.77 \pm 0.09 $ \\

   $\sigma_{\theta}$ &  1.6  &  $ 0.71 \pm 0.01 $    & $ 0.57 \pm 0.01 $   &  $0.45 \pm 0.01 $   & $ 0.60 \pm 0.01 $ \\
   $\sigma_{\phi}$   &  1.6  &  $ 0.80 \pm 0.01 $    & $ 0.78 \pm 0.01 $   &  $0.80 \pm 0.02 $   & $ 2.32 \pm 0.04 $ \\

   $\sigma_{\theta}$ & 25.6  &  $ 0.26 \pm 0.01 $    & $ 0.32 \pm 0.01 $   &  $0.19 \pm 0.01 $   & $ 0.31 \pm 0.03 $  \\
   $\sigma_{\phi}$   & 25.6  &  $ 0.31 \pm 0.01 $    & $ 0.29 \pm 0.01 $   &  $0.33 \pm 0.02 $   & $ 1.04 \pm 0.03 $  \\
   
\hline
   
\end{tabular}
\caption{Sub-sample of values for $\theta$ and $\phi$ resolutions for photons in ILD00. Resolution units are mrad}
\end{center}
\end{table}



With these procedures the pull distributions of the measurements show that we manage to 
model the energy and angular resolutions reasonably well. 

\subsection{Multiple $\pi^0$ Tests using $\pi^0$'s from $Z^0$ Decay} 

The full simulation studies of $Z^0$ events are based on simulating the multiple prompt-$\pi^0$ 
component of 50,000 $Z^0$ events allowing the simulation to decay the $\pi^0$'s. 
We do not at present simulate the rest of the event (consisting mainly of charged hadrons and neutral hadrons).
We require that photons are detected in a fiducial region of $|\cos{\theta}| < 0.978$ in order to be considered for fitting.
In order to restrict to events where our current calorimeter based photon reconstruction estimates work, we also remove events where 
photon conversion ($\gamma \rightarrow \mathrm{e}^{+}\mathrm{e}^{-}$) 
or Dalitz decays ($\pi^{0} \rightarrow \mathrm{e}^{+} \mathrm{e}^{-}\gamma  $) have 
occurred in the tracker by 
scanning for such decays using the Monte Carlo information.
We expect that such events can be used in the future. 
Additionally, there are regions of the detector where the reconstruction 
software does not record the Monte Carlo particle that 
deposited energy (e.g. regions near the beam pipe).
Events where Monte Carlo particles could not be recovered are not considered in the analysis. 

%
%
%

\subsection{Energy Sum Estimates}

In the full simulation of $\pi^{0}$'s there are complicating factors 
that must be dealt with. One of these is that especially 
low energy photons may not be detected or reconstructed by the software 
due to very low energy deposits or too few cells being hit.
The overall reconstructed energy therefore usually underestimates 
the generator energy, and sometimes only one of the two photons from a $\pi^0$ 
is detected. So we find it appropriate to use 
modified definitions of ``generator energy'' in some of our performance estimates.

\begin{table}[h]
\begin{center}
\begin{tabular}{ c l c c }
\hline
  Symbol & Definition & $<E_{\mathrm{tot}}>$ & $<E_{\mathrm{tot}^{'}}>$ \\
\hline
  $E_{g0}$ & Energy from all $\pi^{0}$'s in the event                & 21.2 & 18.1 \\
  $E_{g1}$ & Energy from reconstructed photons                       & 20.7 & 17.7 \\
  $E_{g2}$ & Energy from $\pi^{0}$'s with both photons reconstructed & 17.7 & 15.6 \\
\hline
\multicolumn{2}{c}{$E_{g0} > E_{g1} > E_{g2}$} \\
\end{tabular}
\caption{Definitions of generator energy sums used for analysis of $Z^0$ events. 
Also given are the mean energy sums in GeV for $Z^0$ events, $<E_{\mathrm{tot}}>$, 
and the same quantity after removal of conversions and Dalitz decays, $<E_{\mathrm{tot}^{'}}>$. }
\end{center}
\end{table}

\subsection{Using Truth Information}

First studies used Monte Carlo ``truth'' information to pair up photons properly. 
This should indicate the best possible performance of the
mass-constrained fitting procedure.  Identification of which Monte Carlo particle 
is associated with reconstructed particles from the software
was done using energy deposits in the detector.  Since overlap of energy deposits does occur, 
each cluster was associated with the Monte Carlo particle most responsible for the cluster according to energy.

The process of matching daughter photons with parent $\pi^{0}$'s uses the Monte Carlo decay tree information. 
When a single Monte Carlo particle
was assigned to more than one cluster (due to cluster splitting in the reconstruction software), 
the highest energy cluster was preferred for the matches. 
Additionally, to be consistent with how the matching problem is approached when not using 
truth information, after kinematic fits are performed, matches are cut if the fit probability is less than 1\% 
and the energies of photons that are not matched to their parent $\pi^0$ is estimated using the 
standard calorimetric measurement.


\subsubsection{91.2 GeV $Z^{0}$ Performance}

The event selection yielded\footnote{Note that with an average prompt $\pi^0$ multiplicity 
of around 9.4, even with a relatively small probability per photon of converting 
in the detector material, the overall efficiency for no conversion is small} 11,568 $Z^{0}$ events 
at 91.2 GeV and standard 
reconstruction resulted in a Gaussian fitted $\alpha_{\mathrm{reco}} = 17.9 \pm 0.1\%$ 
for the prompt electromagnetic energy sum 
using generator energy $E_{g0}$ described above with a bias of $ (-7.3 \pm 0.2\%) /\sqrt{E}$. 
The $\alpha_{\mathrm{reco}}$ value and the bias 
was estimated from a Gaussian fit within $\pm$3 rms 
units of the observed distribution 
of the stochastically scaled energy residual, $(E - E_{g})/\sqrt{E_{g}}$. 
This same procedure is used for all subsequent resolution and bias estimates, with the 
fitted $\sigma$ giving the estimated resolution 
parameter, $\alpha$, and the fitted mean giving the bias in units of $1/\sqrt{E}$. 
There is a significant bias in the measured energy due to missing photons using the $E_{g0}$ estimator 
and so we have focussed more on the $E_{g1}$ and $E_{g2}$ estimators in the following studies which give 
similar resolutions on measured energy but a much smaller bias of around $(-1.0 \pm 0.2\%) / \sqrt{E}$.
 
The matching process successfully 
reconstructed 75\% of the generator $\pi^{0}$ energy and improved the energy 
resolution to $\alpha_\mathrm{{fit}} = 13.5 \pm 0.2\%$ (using $E_{g1}$). 
This fraction is primarily due to missing low energy photons
in the final reconstruction, which in turn results 
in lone unmatched photons. 
But the large impact of undetected photons becomes apparent 
if one uses $E_{g2}$ so that only $\pi^{0}$'s with the potential for 
reconstruction are considered. 
This results in $\alpha_\mathrm{{fit}} = 12.2 \pm 0.2\%$.
The improvement ratio $\alpha_\mathrm{{fit}} / \alpha_\mathrm{{reco}} = 0.68$ is 
similar to the $Z^{0}$ toy Monte-Carlo studies for 
angular resolutions of 0.5 and 1.0 mrad.

\subsection{Matching without Truth Information}

Reconstruction of $\pi^{0}$'s without using truth information is done based on 
the $\chi^{2}$ from the kinematic fits. The process involves five main steps:

\begin{enumerate}
  \item Assign all identified photons and the misidentified photons labelled ``neutrons'' as candidate photons.
  \item Perform mass-constrained fits on every pair of candidate photons and generate a list of candidate $\pi^{0}$'s.
  \item Remove all candidate $\pi^{0}$'s where the fit probability is less than 1\%.
  \item Select the ``best'' combination of $\pi^{0}$'s such that each photon is used at most once.
  \item The energy estimate of candidate photons that are not assigned to any $\pi^0$ in the solution 
reverts to the standard calorimetric estimate.
\end{enumerate}

\subsubsection{Scoring with $\chi^{2}$}

There are many possible ways to score a combination of $\pi^{0}$'s to determine which is ``best''. 
The primary method we use is to sum the $\chi^{2}$ from each fit in the solution to obtain an overall global fit probability for each solution. We then select the most probable solution that maximizes the total number of $\pi^{0}$'s reconstructed.  Below is a table that illustrates 
this selection method:

\begin{center}
\begin{tabular}{ c c c c }
  Soln & Fits & $\chi^{2}/\mathrm{dof}$ & Prob \\
\hline
  a    &  6   &  5/6 = 0.83   & 0.544 \\
  b    &  7   &  8.2/7 = 1.17 & 0.315 \\
  c    &  7   &  14/7 = 2     & 0.051 \\
\end{tabular}
\end{center}

In this example, the maximum number of $\pi^{0}$'s found was seven, so we eliminate option $a$. 
Between options $b$ and $c$, the best $\chi^{2}$
is $b$.

\subsubsection{Combinatorics}

The number of potential solutions to consider after all candidate $\pi^{0}$'s have been generated at step 2 above can be
very large. For $n$ photons with $n$ being even, the number of solutions 
consisting of $n/2$ $\pi^0$'s goes as $(n-1)(n-3)(n-5)...1$.
Fortunately the 1\% fit probability cut usually dramatically reduces this, 
but the issue of scale still exists. Even with 
just 10 photons, typical of say a jet with 5 $\pi^0$'s, this number is 945.

We have employed methods from graph theory to approach this problem.  The problem of matching
photons to parent pions can be modeled using a graph. 
Each vertex of the graph represents a photon and each edge
represents the potential pairing between photons.  We then assign a weight 
to each edge according to the chi-squared of the mass-constrained fit. 
This then becomes a problem of selecting edges such that we use each vertex (photon)
at most once and we minimize the sum of the edge weights (min $\chi^{2}$).

This problem is very close to an efficiently solved problem in graph 
theory classified as a weighted non-bipartite matching problem in which one tries to find 
a perfect matching which minimizes the sum of the edge weights.
A perfect matching is one where each vertex is used exactly once.
For a perfect matching to exist, the graph must have an even number of vertices.

Clearly a real detector can not be expected to only detect an even 
number of photon candidates.
Therefore, we use a method for converting any weighted graph into one where 
a perfect matching exists~\cite{Schafer}. 
This involves creating a new graph by first duplicating the original, such that 
one now has two identical graphs. Then one
creates new edges
connecting each original vertex with its duplicate.
This extra step allows photons to go unmatched which is a desirable feature given that  
we expect that some photons should not be used to 
reconstruct $\pi^{0}$'s (e.g. due to missing sibling photons).

Once the graph has been constructed, we can then use an efficient method for finding the solution. 
One such method, is the latest implementation of Edmonds' algorithm~\cite{Edmonds} called 
Blossom V, by Kolmogorov~\cite{Kolmogorov} which can solve the graph in polynomial time. 
The worst case performance for a graph with $n$ vertices and $m$ edges is $O(n^3 \, m)$ but 
on average is much faster. This appears to be well suited to the problem as defined but is not 
necessarily convenient for systematically exploring alternative solutions.


\subsection{91.2 GeV $Z^{0}$ Performance}

As with using truth information, the standard reconstruction yielded 11,568 events 
and an energy resolution of $\alpha_{\mathrm{reco}} = 17.9 \pm 0.1\%$.
The above matching procedure was able to 
correctly reconstruct 51\% of the generator $\pi^{0}$'s which 
represented 68\% of the total
$\pi^{0}$ energy per event ($E_{g1}$). 
This resulted in an effective energy 
resolution of $\alpha_{\mathrm{fit}} = 14.6 \pm 0.2 \%$ and 
a small bias of $(-2.5 \pm 0.2 \%)/\sqrt{E}$.  

When one looks at $\pi^{0}$ energy where both photons are 
reconstructed by the software, the improvement to $\alpha_{\mathrm{fit}}$ then
becomes $13.9 \pm 0.2 \%$ with a small bias of $ (-2.8 \pm 0.2 \%)/\sqrt{E}$.  
In this case, $78\%$ of the $\pi^{0}$ energy 
that had the potential for fitting (i.e. $E_{g2}$)
was fit correctly and $\alpha_{\mathrm{fit}} / \alpha_{\mathrm{reco}} = 0.77$.

\begin{table}[h]
\begin{center}
\begin{tabular}{ l c c c c c c c c c }
Generator & 
\multicolumn{2}{c}{Standard} & 
\multicolumn{2}{c}{$\pi^{0}$ Reconstruction} & 
\multicolumn{2}{c}{Corrected} & 
\multicolumn{2}{c}{Truth} \\
Energy & $\alpha_\mathrm{{reco}}$  & bias &  $\alpha_\mathrm{{fit}}$ & bias & $\alpha_\mathrm{{fit}}$ & bias  &  $\alpha_\mathrm{{fit}}$ & bias \\
\hline
$E_{g0}$  &  17.9 & -7.3 & 15.4 & -8.9 & 14.6 &  -7.0 & 13.9 & -4.7 \\
$E_{g1}$  &  18.0 & -0.6 & 14.6 & -2.5 & 14.1 &  -0.6 & 13.5 &  1.8 \\
$E_{g2}$  &  18.0 & -0.8 & 13.9 & -2.8 & 13.0 &  -0.8 & 12.2 &  1.8 \\

\end{tabular}
\caption{Summary of mass-constrained fit procedure results in \% applied to multiple 
$\pi^{0}$'s from $Z^{0} \rightarrow q\bar q$.  Errors on 
the $\alpha$ and bias values are approximately 0.2\%}
\label{tab:results}
\end{center}
\end{table}

\subsection{Effects of Incorrect Photon Pairings}

The $\pi^{0}$ reconstruction process occasionally pairs photons incorrectly and 
thus generates $\pi^{0}$'s which have been fitted with an inappropriate 
mass constraint.
In the above example, while 51\% of
$\pi^{0}$'s were reconstructed, an additional 19\% were incorrectly identified.
The effects of these errors can be seen by removing incorrect $\pi^{0}$'s and replacing 
with the original photons that were measured with the calorimeter and then comparing 
the final results. In the case of 91.2 GeV $Z^{0}$'s these ``corrected'' events 
improve the effective energy resolution to $14.1 \pm 0.2\%$ with a negligible 
bias (using $E_{g1}$). 
When using the $E_{g2}$ estimator, and correcting for incorrect pairings, the resolution 
decreases to $\alpha_{\mathrm{fit}} = 13.0 \pm 0.2 \%$ and the bias is also negligible.
Overall this indicates that incorrect pairings do negatively influence the 
resolution performance in a significant manner 
and that there is some scope for further improvement if one can reduce the error rate.
The bias from incorrect pairings is small and can likely be calibrated out.

\subsection{Results Summary}
A summary of the results comparing the resolution and bias estimates from the different estimators 
and pairing algorithms are given in Table~\ref{tab:results}. Individual distributions for the different 
pairing algorithms are shown in Figures 1-4 for the generator energy sum estimates which 
are corrected for undetected photons. It is clear from all the distributions 
that the $\alpha$ parameter determined from the Gaussian fits does not fully account for 
the observed peakedness of the distributions based on mass-constrained fits. 
Such behavior results from $\pi^0$'s with small $|\cos{\theta^*}|$ 
being measured with much better energy resolution after the fit than the average fitted $\pi^0$.

\begin{figure}
\begin{minipage}[t]{\linewidth}
\centering
\includegraphics[scale=.7]{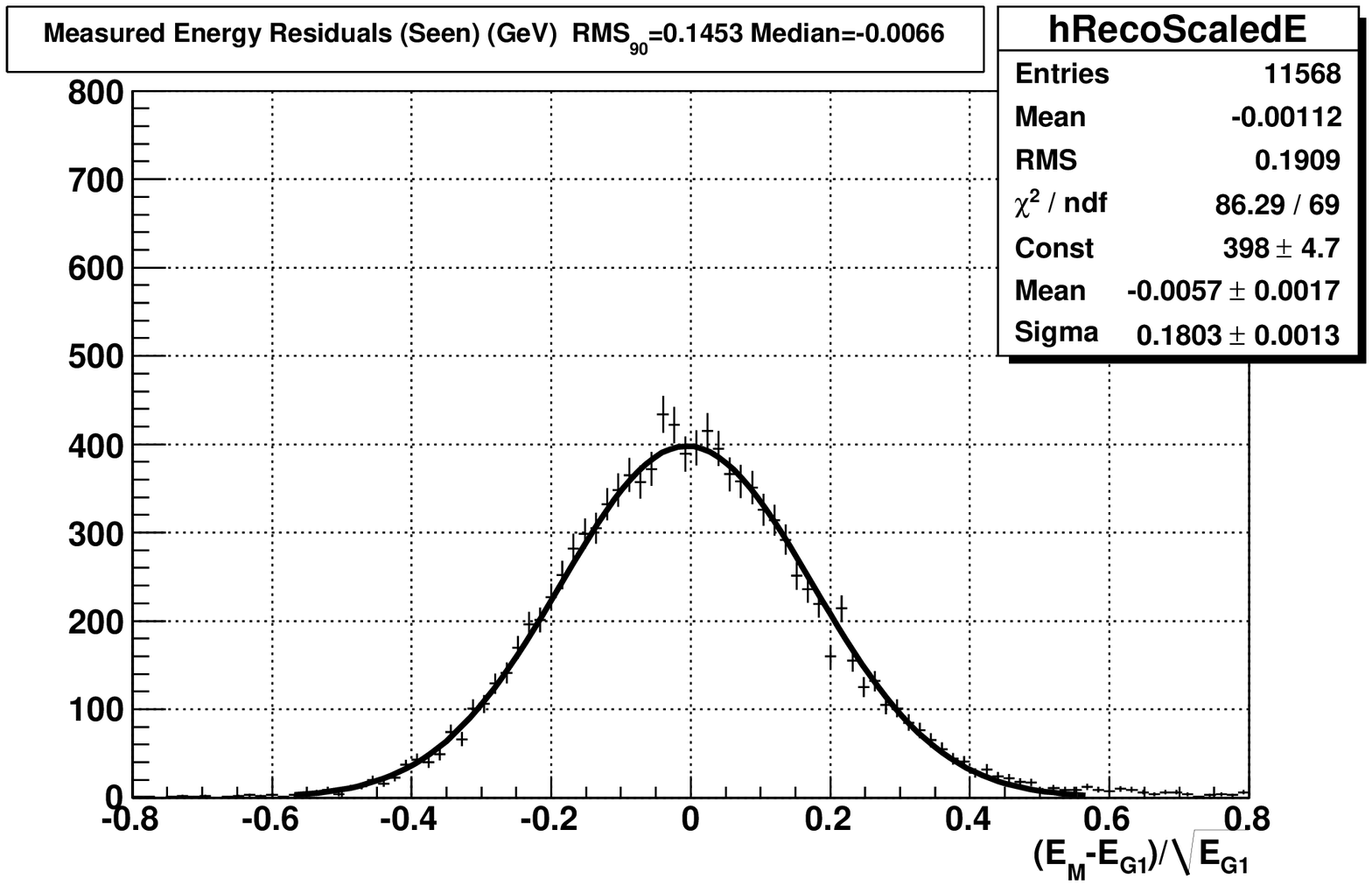}
\vspace{0.2cm}
\includegraphics[scale=.7]{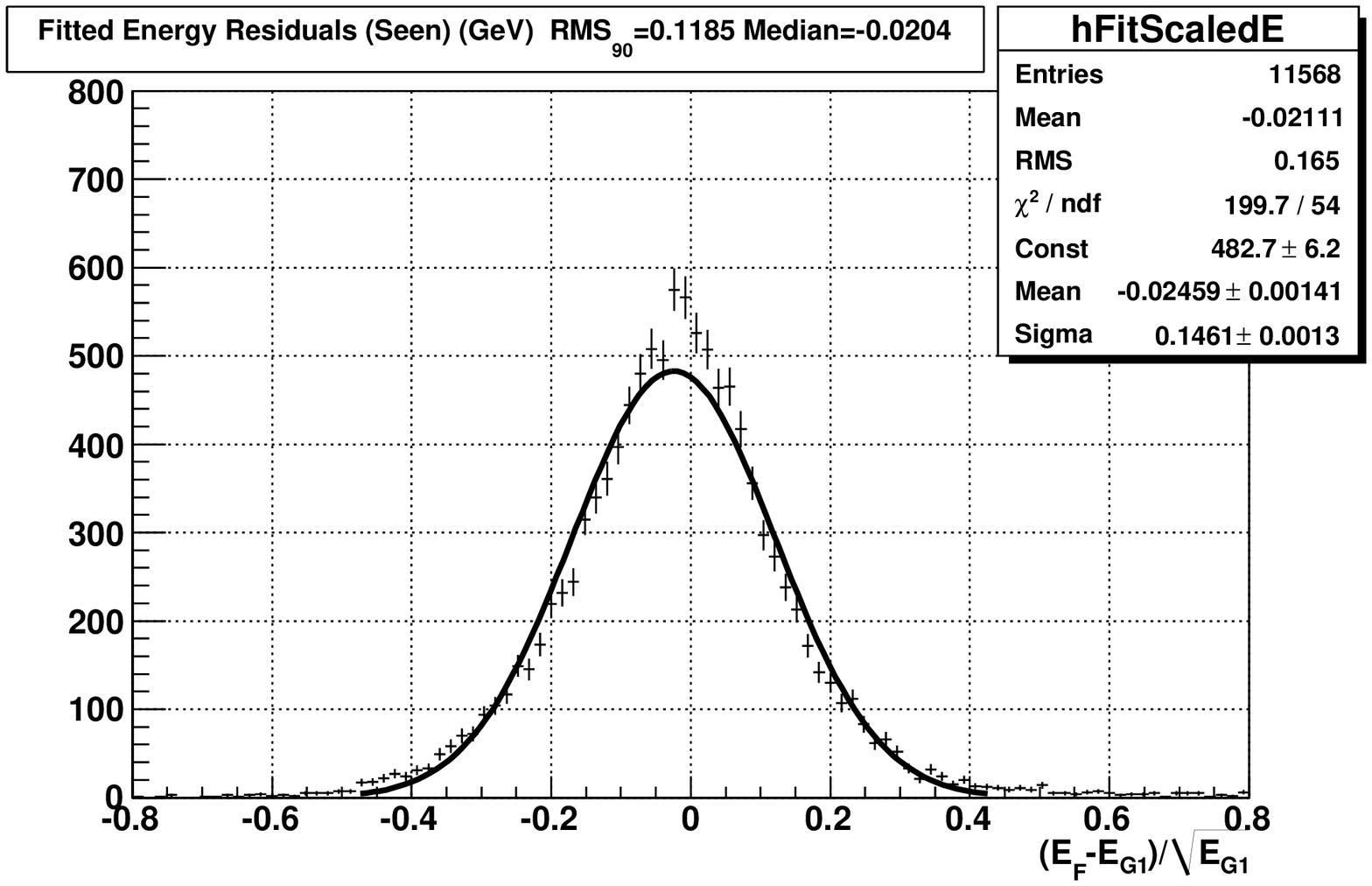}
\end{minipage}
\caption{Measured stochastically scaled energy residuals $(E - E_{g})/\sqrt{E_{g}}$ 
for different pairing algorithms using $E_{g1}$ as the reference energy.
The top panel shows standard reconstruction and the bottom panel shows the 
realistic matching algorithm.
}
\end{figure}

\begin{figure}
\begin{minipage}[t]{\linewidth}
\centering
\includegraphics[scale=.7]{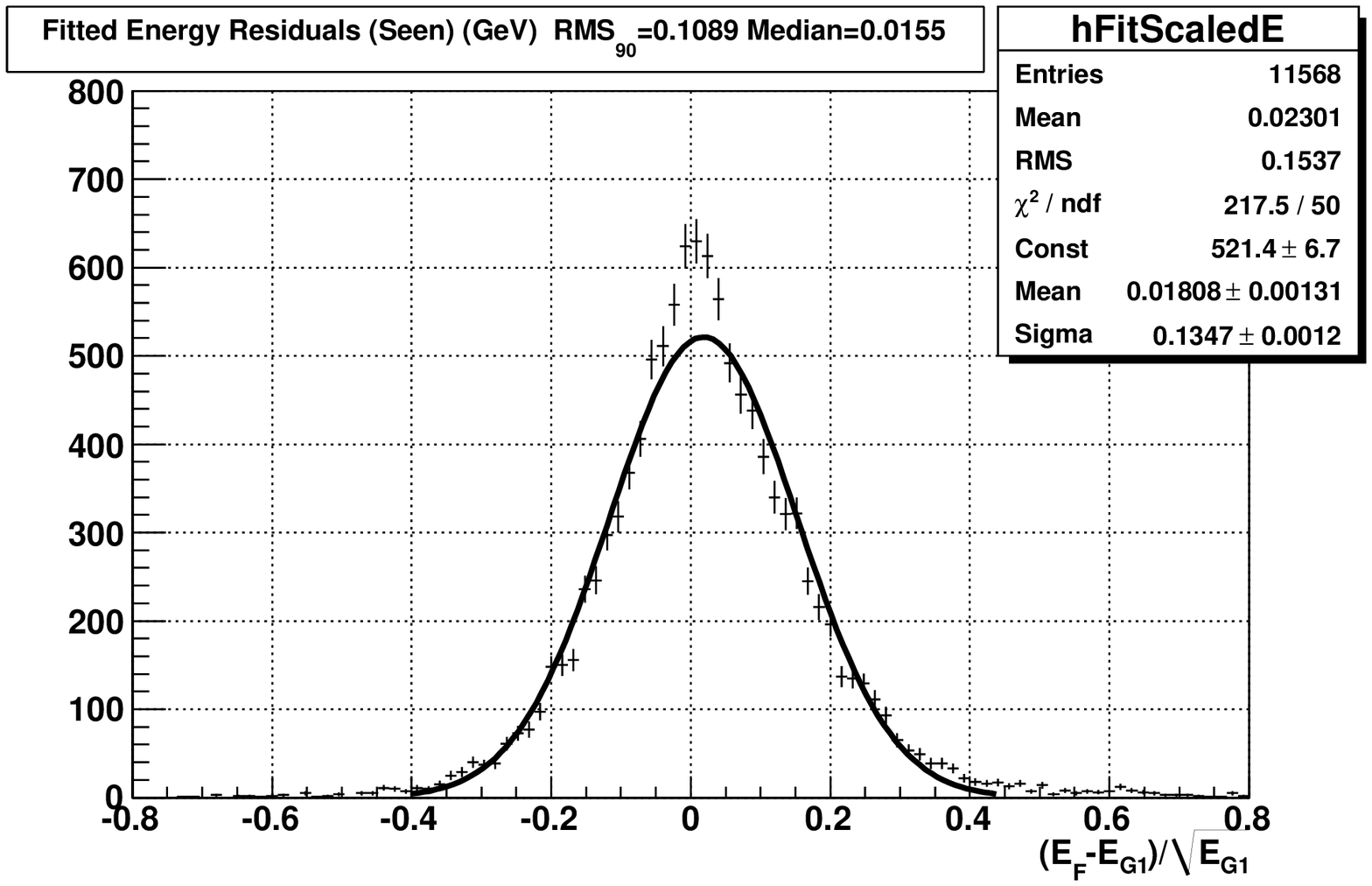}
\vspace{0.2cm}
\includegraphics[scale=.7]{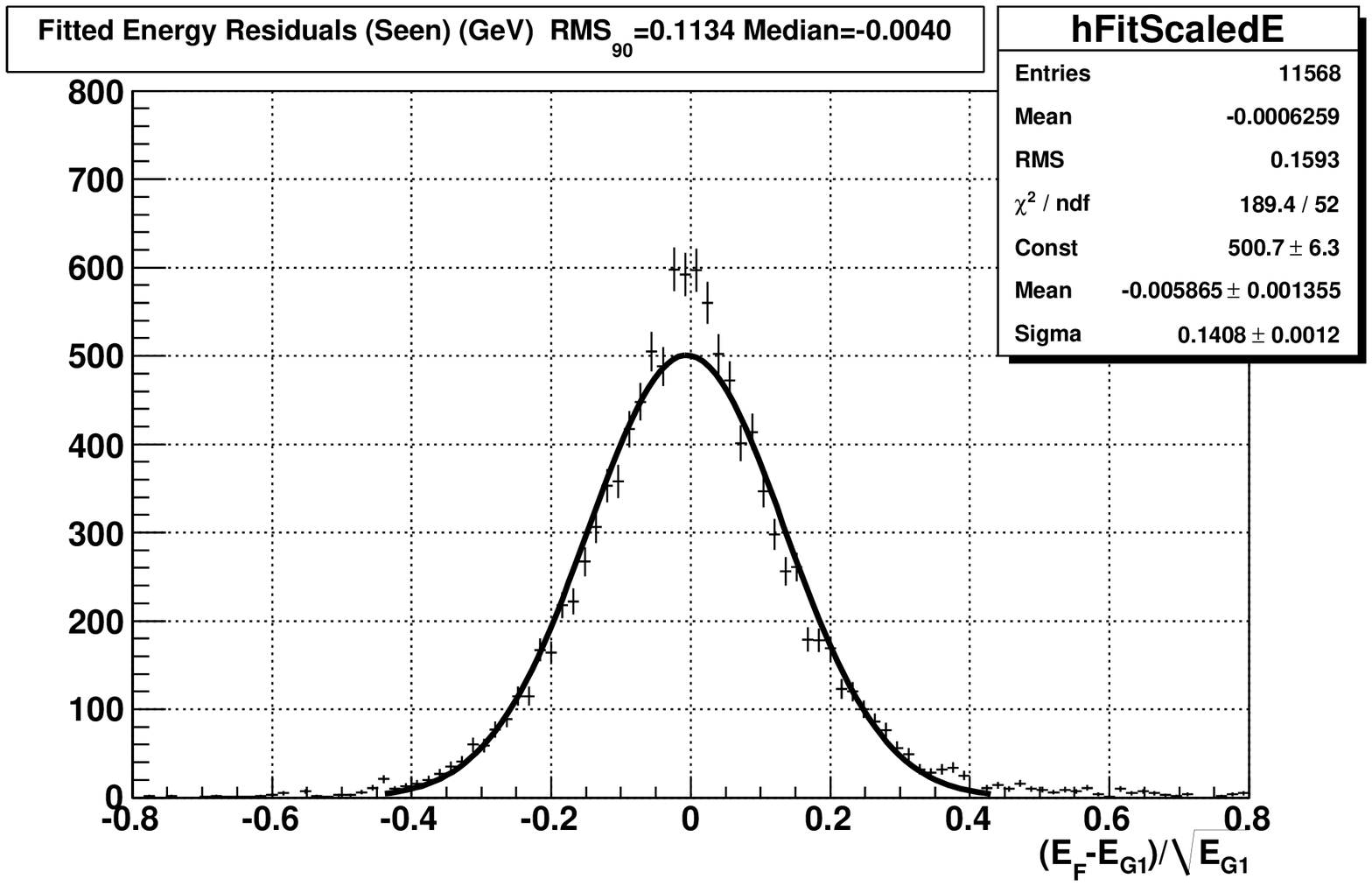}
\end{minipage}
\caption{Measured stochastically scaled energy residuals $(E - E_{g})/\sqrt{E_{g}}$ 
for different pairing algorithms using $E_{g1}$ as the reference energy.
The top panel shows the cheated version and the bottom panel shows the 
realistic matching algorithm but with incorrect pairings replaced.
}
\end{figure}

\begin{figure}
\begin{minipage}[t]{\linewidth}
\centering
\includegraphics[scale=.7]{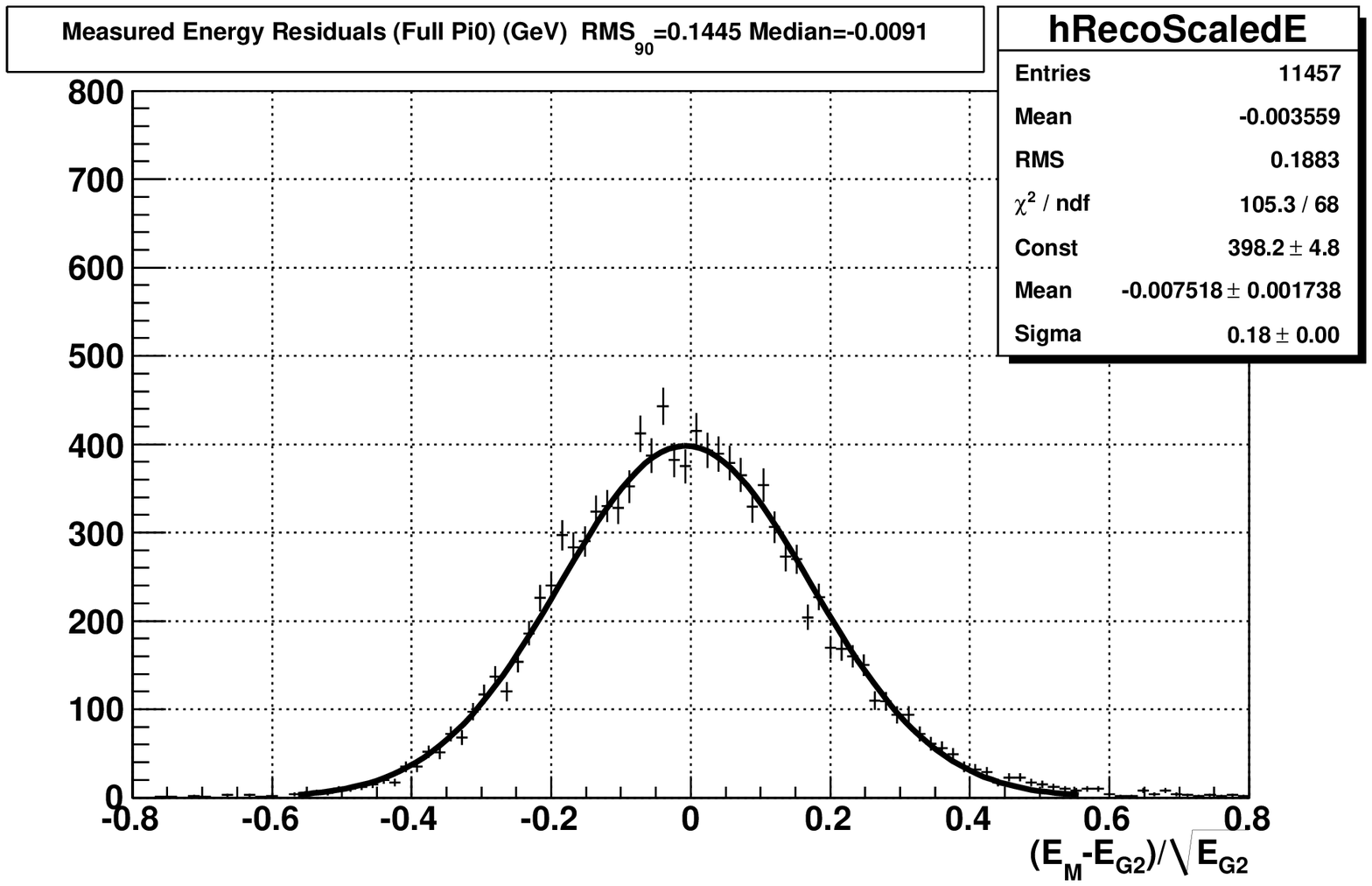}
\vspace{0.2cm}
\includegraphics[scale=.7]{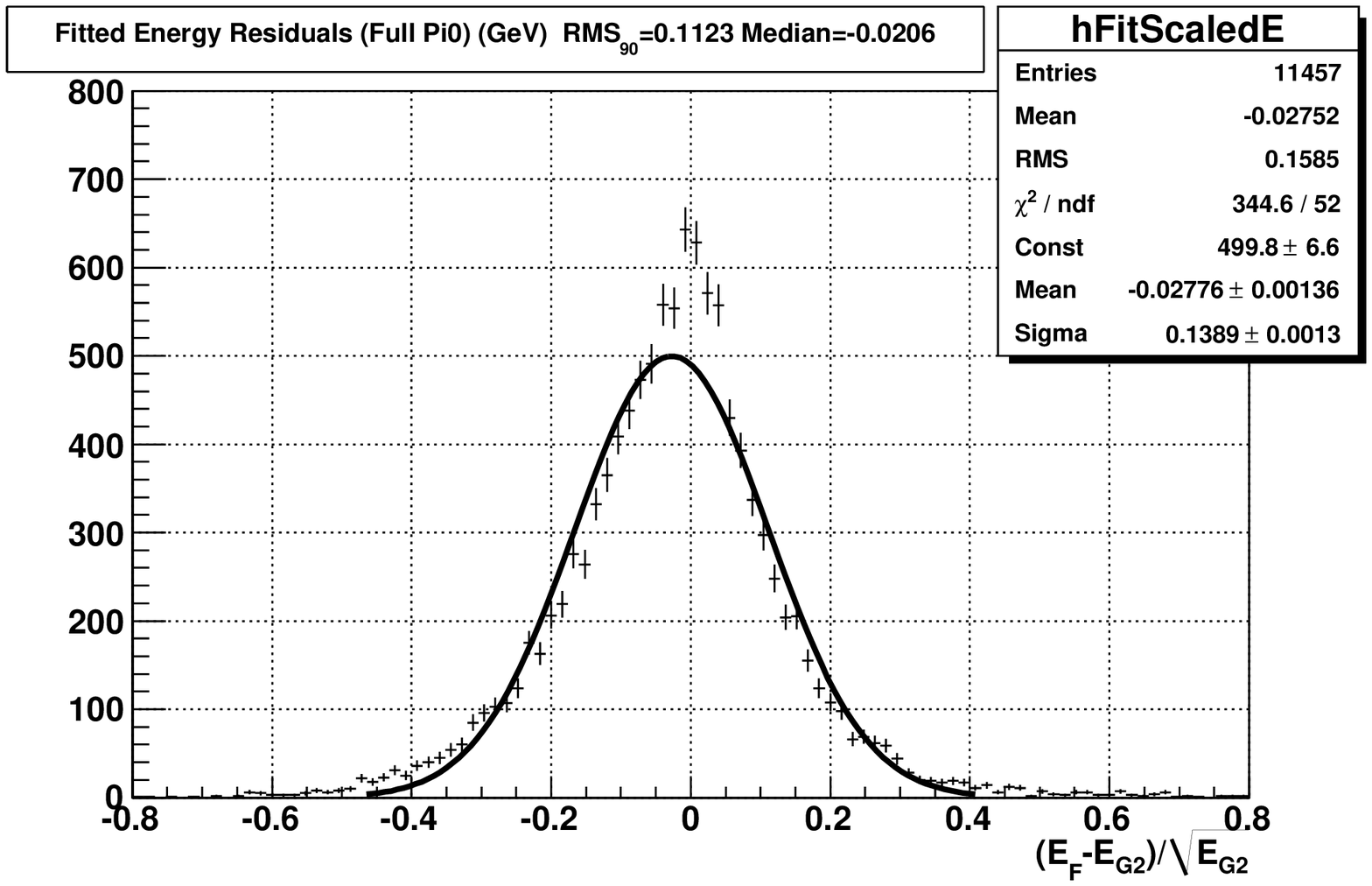}
\end{minipage}
\caption{Measured stochastically scaled energy residuals $(E - E_{g})/\sqrt{E_{g}}$ 
for different pairing algorithms using $E_{g2}$ as the reference energy.
The top panel shows standard reconstruction and the bottom panel shows the 
realistic matching algorithm.
}
\end{figure}

\begin{figure}
\begin{minipage}[t]{\linewidth}
\centering
\includegraphics[scale=.7]{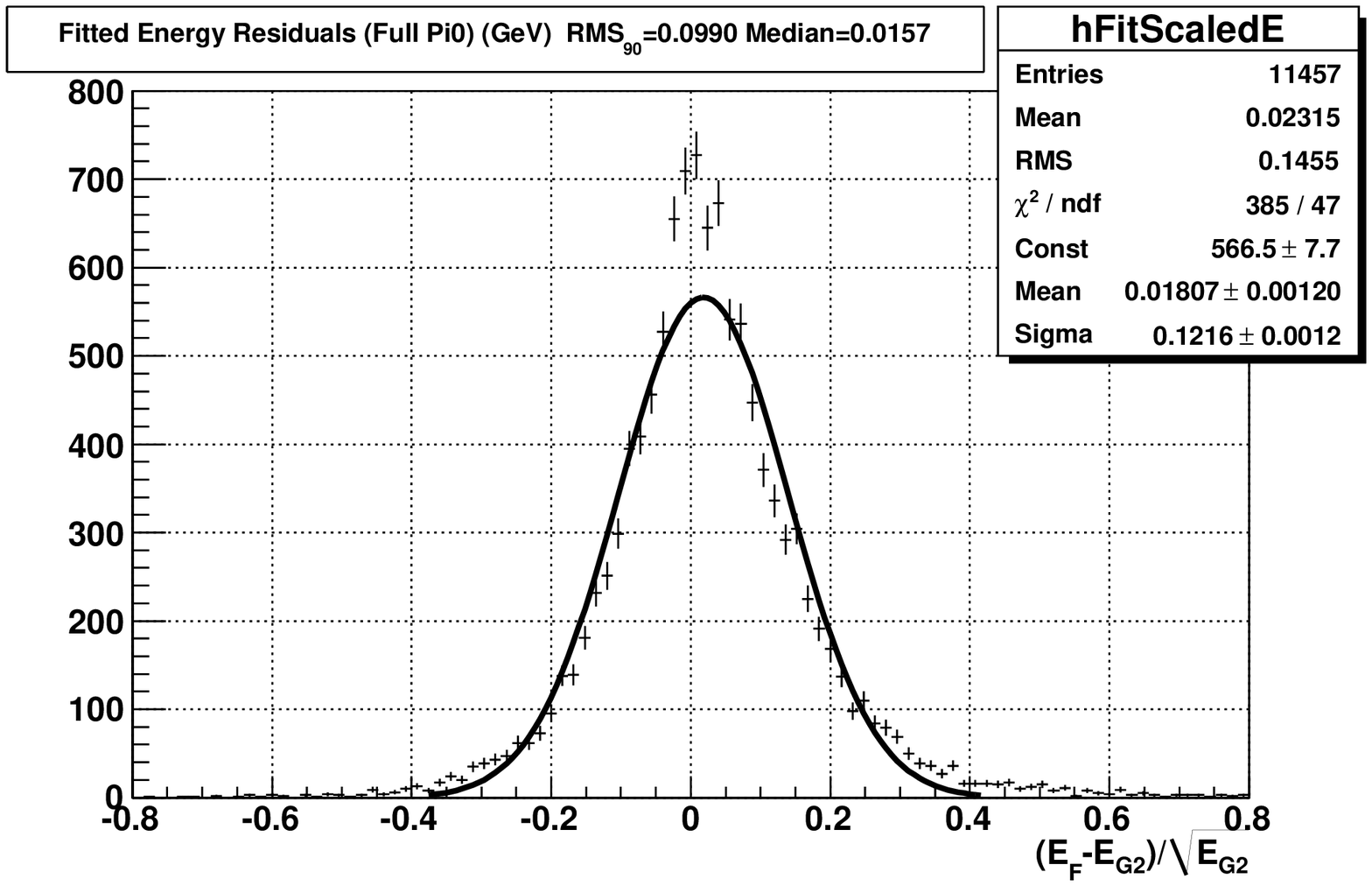}
\vspace{0.2cm}
\includegraphics[scale=.7]{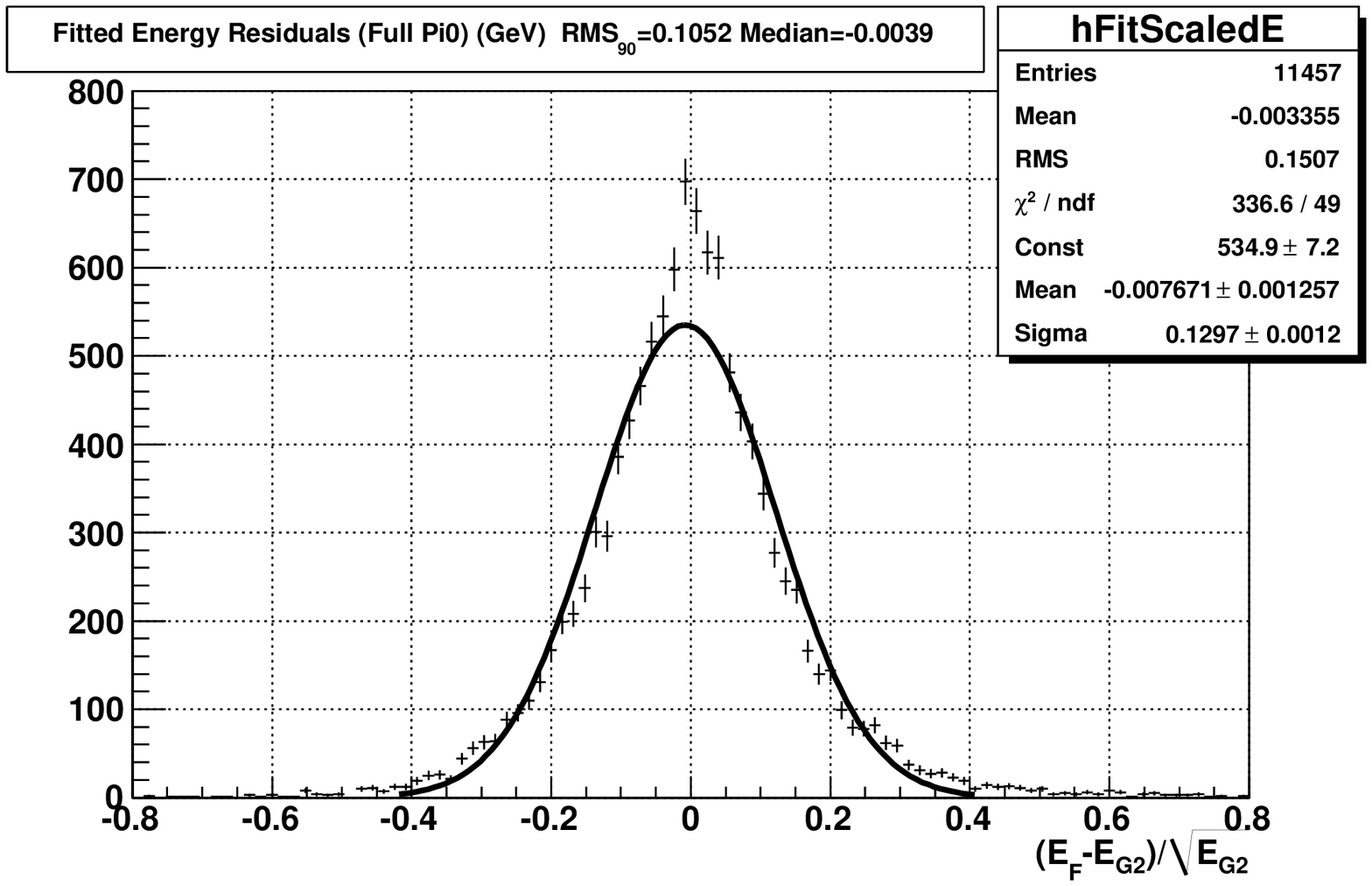}
\end{minipage}
\caption{Measured stochastically scaled energy residuals $(E - E_{g})/\sqrt{E_{g}}$ 
for different pairing algorithms using $E_{g2}$ as the reference energy.
The top panel shows the cheated version and the bottom panel shows the 
realistic matching algorithm but with incorrect pairings replaced.
}
\end{figure}

\subsection{Tuning the Algorithm}

There are three main parameters that currently can be tuned.
$p_\mathrm{{min}}$ is the minimum fit probability necessary to be considered part of a final match solution.  The single photon chi-squared, $\chi^{2}_{\gamma}$ represents the
edge weight between photons cloned during the graph modification phase that guarantees that a perfect matching exists. It essentially allows for
photons to go unmatched in the final solution. $E_{\gamma}^\mathrm{{min}}$ is the minimum photon energy that is required 
for a photon to be considered available to be matched. 
In the case of 91.2 GeV $Z^{0}$'s we found optimal values 
to be: $p_{\mathrm{min}} = 0.01$, $\chi^{2}_{\gamma} = 6.6348$ (equivalent 
to 1\% fit probability), and $E_{\gamma}^{\mathrm{min}} = 50$ MeV.
Any significant deviations from these values resulted in a worsening of performance.  It is likely that one or 
more of these depends on characteristics of the events (e.g. number and energy of particles) and 
is an issue we are currently working on.

\section{Outlook}

Efficient reconstruction of fitted $\pi^{0}$'s depends significantly on how 
well individual photons are reconstructed. There are a number of areas 
where the photon reconstruction efficiency could be improved either 
by algorithmic improvements or potentially by improving the detector design.

\begin{itemize}
\item{At high energies, photon energy deposits tend to be split into several sub-clusters}
\item{At energies below about 180 MeV the efficiency to reconstruct photons degrades reaching 50\% 
at 100 MeV, and lower still for lower energies}
\item{Photons which convert in the tracker are currently not reconstructed well and 
should be amenable to reconstruction with high efficiency and excellent resolution}
\end{itemize}

The loss of low energy photons is of particular concern because it leads 
to cases where an energetic photon from a $\pi^0$ may more 
easily be paired with the wrong photon. 
Cases where one of the photons from a $\pi^0$ converts in the tracker and Dalitz decays 
should lead to a significant improvement in the overall energy resolution on the $\pi^0$ through 
mass-constrained fitting. 
Other directions for improvement include improving upon the photon 
position reconstruction, the use of the information from the fitted 
uncertainties in assigning jet-specific energy resolution uncertainties, and 
the assessment of performance for events with higher energy and particle multiplicity.

\section{Conclusions}

We have shown that with a  
full simulation of the ILD detector response to multiple $\pi^0$'s 
that it is feasible to 
improve substantially the prompt electromagnetic component of jet energy 
resolution using mass-constrained fits.
For events where both photons of the $\pi^0$ are detected, 
the resulting solutions lead to an improvement 
in the $\pi^{0}$ component of the event energy resolution 
for 91.2 GeV $Z^{0}$ events 
from $18.0\%/\sqrt{E}$ to $13.9\%/\sqrt{E}$ using 
the ILD00 detector and its reconstruction algorithms.
This can be compared to a maximum potential improvement to $12.2\%/\sqrt{E}$ 
if all photon pairs are matched correctly using the current photon reconstruction.

\section{Acknowledgements}
We express our thanks to various people 
who have helped in various capacities associated directly with this work. 
Jeremy Martin for pointing us towards the matching literature. 
Norman Graf for his continued interest and support. 
John Marshall for help with interfacing to PandoraPFANew. 
Benno and Jenny List for their support of the MarlinKinFit package. 
Vladimir Kolmogorov for help with our questions on Blossom V.

%


\begin{footnotesize}


\end{footnotesize}



\begin{thebibliography}{99}


\bibitem{Brient} 
  J.~-C.~Brient and H.~Videau,
  {\it The Calorimetry at the future $e^{+} e^{-}$ linear collider},
  eConf C {\bf 010630}, E3047 (2001)
  [hep-ex/0202004].

\bibitem{Morgunov} 
  V.~L.~Morgunov,
  {\it Energy flow method for multi-jet effective mass reconstruction in the highly granulated TESLA calorimeter},
  eConf C {\bf 010630}, E3041 (2001).

\bibitem{Thomson} 
  M.~A.~Thomson,
  {\it Particle Flow Calorimetry and the PandoraPFA Algorithm},
  Nucl.\ Instrum.\ Meth.\ A {\bf 611}, 25 (2009)
  [arXiv:0907.3577 [physics.ins-det]].

\bibitem{MarlinKinFit}
B.~List and J.~List, {\it MarlinKinFit: An Object Oriented Kinematic Fitting Package}, LC-TOOL-2009-001. 

\bibitem{Graham_OldTalks}
G.~W.~Wilson and B.~van Doren, 
{\it Towards Jet Specific 
Energy Resolution: Investigating $\pi^0$ Kinematic Fits}, 
2009 Linear Collider Workshop of the Americas, Albuquerque, NM, October 2009.

\bibitem{PDG}
K.~Nakamura {\it et~al.}, (Particle Data Group), J. Phys. G 37, 075021 (2010).

\bibitem{LOI}
T.~Abe {\it et al.}  [ILD Concept Group - Linear Collider Collaboration],
{\it The International Large Detector: Letter of Intent},
arXiv:1006.3396 [hep-ex], published as
DESY 2009-87, FERMILAB-PUB-09-682-E, KEK Report 2009-6.

\bibitem{Graham_position_talk}
G.~W.~Wilson, {\it Improving Photon Parameter Estimation Using Shower Fitting: A Status Report}, LCWS11 Proceedings.

\bibitem{Schafer}G.~Sch\"{a}fer, {\it Weighted matchings in general graphs}, Master's thesis, 
Fachbereich Informatik, Universit\"{a}t des Saarlandes, Saarbr\"{u}cken, Germany (2000).

\bibitem{Edmonds}
J.~Edmonds, {\it Paths, Trees and Flowers}, Canadian Journal of Mathematics 17: 449-467.

\bibitem{Kolmogorov}V.~Kolmogorov, {\it Blossom V: A new implementation of a minimum cost perfect matching algorithm}, Mathematical Programming Computation 1 (1): 43-67 (2009). We used version 2.02.



\end{thebibliography}
\end{document}